\documentclass[11pt]{article}
\usepackage{graphicx}
\usepackage{amsmath,amssymb,epsfig,bm}

\newcommand\tr{\mathop{\mathrm{tr}}}

\newcommand{\be}{\begin{eqnarray}}
\newcommand{\ee}{\end{eqnarray}}

\def\lsim{\mathrel{\rlap{\lower3pt\hbox{\hskip1pt$\sim$}}
     \raise1pt\hbox{$<$}}} 
\def\gsim{\mathrel{\rlap{\lower3pt\hbox{\hskip1pt$\sim$}}
     \raise1pt\hbox{$>$}}} 

\def\la{\langle}\def\ra{\rangle}
\def\del{\partial}

\newcommand\Tr{\mathrm{Tr}\,}

\def\bi{\bibitem}
\def\tr{{\rm tr}}

\renewcommand{\thefootnote}{\fnsymbol{footnote}}

\begin{document}

\begin{titlepage}
\begin{flushright}

\end{flushright}
\vskip 1.7in
\begin{center}
{\bf\Large{Subtle Is The Manifestation Of Chiral Symmetry \\
 In Nuclei And Dense Nuclear Matter}}\footnote{To appear in {\it Festschrift for Gerry Brown}, ed. Sabine Lee (World Scientific, Singapore)}
\vskip 0.2cm
\today

\vskip 1.5cm

{Mannque Rho}
\vskip 0.05in
{\small{\textit{Institut de Physique Th\'eorique,  CEA Saclay, 91191 Gif-sur-Yvette C\'edex, France \&\\
Department of Physics, Hanyang University, Seoul 133-791, Korea}}}

\end{center}
\vskip 0.7in
\baselineskip 16pt

\begin{abstract}
The history of how chiral symmetry has entered in nuclear physics, in which Gerry Brown and I have participated from 1970 up to today, is described from my personal viewpoint. The route of development we have traversed together goes from meson exchange currents, to ``little chiral bag," to chiral effective field theory, to ``Brown-Rho scaling" and then to dense matter and chiral phase transition. It has been a great fun and exciting, some right and some wrong in what we have done together, but none that can be dismissed as ``not even wrong." We have found all along that whatever signal there may be for the manifestation of chiral symmetry in nuclear medium, be it at low density in meson exchange currents or at high density approaching the chiral phase transition, is rich in its intricacy and subtlety.
\end{abstract}

\end{titlepage}
\setcounter{footnote}{0}
\renewcommand{\thefootnote}{\arabic{footnote}}
\section{Introduction}
Gerry Brown and I started to ask how chiral symmetry manifests itself in nuclear structure and dense nucleonic matter way back in late 1960's and have been chasing after it ever since. At that early time, quantum chromodynamics was not yet around and hence chiral symmetry was not considered in terms of nearly zero mass quarks. What was available then was a set of current-algebra relations associated with pions as Nambu-Goldstone bosons, and it was agreed that pions must play an important role in nuclear interactions. Gerry, at the time at Nordita in Copenhagen, was looking at the problem in connection with nuclear forces~\cite{GEB1} and I, freshly tenured at Saclay after a year of post-doc there, was more interested in how chiral symmetry could be exploited in nuclear responses to external fields, such as electromagnetic and weak.

In this note I would like to describe the line of development I followed, some in collaboration with him and some with others but strongly influenced by Gerry's thinking.

\section{How it all came about}
With Marc Chemtob, then a thesis student at Saclay working on exchange currents in nuclei, we were asking how the soft-pion theorems associated with the current algebras worked out by particle theorists could be used to constrain the pion-exchange currents which we thought would dominate the processes at not too large momentum transfers.

We were motivated to address this question from conversations with Benjamin Lee who was at that time on his sabbatical stay at Institut des Hautes \'Etudes Scientifiques at Bures-sur-Yvette, which is located very near Saclay, and was on his celebrated work on the renormalization of the sigma model --  a prelude to the renomalization of Yang-Mills gauge theory. These stimulating and fruitful conversations with Ben Lee led us to write in 1969-70 a paper where exchange currents were derived using S-matrix formalism in which the soft-pion theorems were invoked in calculating the longest-range exchange currents~\cite{CR69}.

It turned out that the result obtained in \cite{CR69} was instrumental to Gerry and Dan Olof Riska for their pinning down,  to a few\% accuracy, the long-standing exchange-current contribution in the thermal $p+n\rightarrow d+\gamma$ process~\cite{riska-brown}. This calculation was later confirmed by chiral perturbation theory, an effective field theory of QCD, to which I will return below. This set in motion my long collaboration with Gerry on how chiral symmetry, later identified with an essential ingredient of QCD, figured in not just organizing corrections to nuclear matrix elements of the electroweak currents, but more importantly in playing an indispensable role in the basic structure of the nucleon itself and its vacuum changes in hot and dense background. Along the way, we have come up with the concept of ``Little Chiral Bag," followed by Brown-Rho scaling and their place in nuclear chiral perturbation theory developed in 1990's.  Many diverse phenomena observed in nuclei have been found naturally to fit in the scheme, influencing our later work on dense (and to some extent hot) matter.

When there is a novel concept involved, one would naturally like to check it by singling out experimentally a distinctive signal -- a ``smoking-gun" -- for the effect. A naive thinking would be that if the proton mass of $\sim 1$ GeV were generated (mostly if not entirely) by spontaneous breaking of chiral symmetry as widely believed, with the quark mass being near zero, then one should be able to ``unbreak" the symmetry in the laboratory by heating the system to high temperature or compressing it to high density at which the symmetry breakdown is supposed to take place. Such experiments have indeed been performed. What has come out of such searches turns out, however, to be unclear, both unsettling and intriguing. Indeed experiments done at the SPS/CERN and the RHIC/Brookhaven in heavy ion collisions have failed so far to isolate a ``smoking-gun" signal for the effect. At first thought, this seems disappointing. At second thought, however, it is intriguing that chiral symmetry -- which must be broken in nuclear systems as indicated, though indirectly, in a variety of nuclear processes -- could hide from detection. In this note contributing to Gerry's 85th birthday celebration, I would like to describe how the chiral symmetry that shows up unambiguously in exchange currents -- and also in various other ways -- can hide from view in those experiments specifically looking for a direct evidence. It is significant that the same old pions that we started to take seriously in what we were doing a long time ago turn out to take up a surprising new role today on par with quarks and gluons in a modern effective theory of QCD~\cite{weinberg10}. This development seems to resuscitate the old idea of ``BR scaling" in the original form Gerry and I proposed in 1991 although there are recent indications that this form should be modified at high density.
\section{From soft-pion theorems to chiral perturbation theory}
One of the pressing questions in nuclear physics in 1960's was how meson degrees of freedom figured in nuclear interactions. While it was generally accepted that the pion entered in giving the nuclear forces, it was not very well understood in what way. The question then was how one can ``see" in a form of direct signal its role in nuclear dynamics. To put the question more specifically, consider first two-nucleon systems in strong interactions below the pion production threshold, the pion being the lowest-mass meson with mass $m_\pi\approx 140$ MeV. Given a well-determined two-nucleon potential $V_2$, two-nucleon properties, such as scattering, binding energy etc. can be computed more or less accurately, with no explicit intervention of other degrees of freedom than the nucleons. However, the situation is changed when an external field is applied to the system, which is given by the matrix element
\be
M=\la f|{\cal M}|i\ra\label{me}
\ee
where $i$ and $f$ are, respectively, the initial and final wave functions of the states involved and ${\cal M}$ is the appropriate probe operator, weak or EM in our case. Expressed in terms of the nucleon variables, the operator ${\cal M}$ will consist of the sum of one body operators ${\cal M}_1+{\cal M}_2$ and a two-body operator ${\cal M}_{12}$. ``Accurate" wave functions are given if the potential $V_2$ is accurately known. Assuming symmetries involved in the process, a fully ``realistic" potential can be determined provided there are enough experimental data points available to fix the parameters figuring in the potential. Such a potential is also referred to as ``sophisticated potential"~\footnote{By ``realistic" or ``accurate" or ``sophisticated" two-nucleon potential, one means a potential that is fit to a large number of scattering data up to lab energy $\lsim 300$ MeV with $\chi^2$/datum $\lsim 1$. The present status of chiral perturbation calculation of the NN potential at the highest chiral order reached -- which is next-to-next-to-next leading order or $N^3LO$ -- is that one has to go still higher in chiral order to reach the accuracy of the most ``sophisticated" potential. It is unlikely that such a calculation is fully doable within the foreseeable future.}.

Now suppose the matrix element of the operator ${\cal M}$ between single nucleons is known from on-shell experiments. This gives the single-particle operator. Then one may take as the first approximation for the two-nucleon system the sum ${\cal M}_1+{\cal M}_2$. In this case, $M$ will be given entirely by on-shell nucleon quantities. This leads to what is called ``impulse approximation." Given accurate wave functions, {\em any} deviation from the experimental value must then be attributed to a two-body contribution that comes from other degrees of freedom than the nucleons that figure in the interactions. In terms of the Yukawa meson-exchange picture for nuclear forces, this corresponds to the ``meson-exchange-current" contribution to $M$.

The $\sim 10\%$ deviation of the impulse-approximation calculation from the experimental cross section in the radiative capture
\be
n+p\rightarrow d+\gamma\label{npcapture}
\ee
clearly indicated the presence of meson degrees of freedom contributing to the current. How to calculate this correction had been a long-standing problem until early 1970's.

Given a reliable field theory technique to calculate the potential that fits the scattering data, the same theory should enable one to compute the exchange currents in consistency with the potential. But there was no such theory available then. The wave functions are what Nature is supposed to be giving to the extent that they are extracted from experiments but how to compute the exchange currents compatible with the wave functions was unknown. In \cite{CR69}, it was proposed that the currents associated with the operator ${\cal M}$ be calculated via an S-matrix method, involving necessarily on-shell quantities, constrained by the assumed symmetries of the current. For the EM current, there is the vector current conservation, and for the weak current, both the vector-current and axial-vector-current conservations. Such conservation laws applied to the total current matrix elements give certain relations between two-body currents and two-body potentials. A well-known example is the Siegert theorem for the vector current. One can write down an analogous relation for the axial current, but here the exchange current is not directly related to the NN potential.

The question addressed in \cite{CR69} was how best to write down the total current compatible, if not fully consistent in the sense of the modern chiral counting as is done nowadays, with the potential used to get the wave functions. The key point of \cite{CR69} was that the longest-range exchange current for low-momentum-transfer processes could be given reliably by a pion exchange and that the pion exchange be constrained by the low-energy theorems. Indeed the exchange currents so constructed -- with an additional correction from the $\Delta$ resonance which turns out to be higher order in the chiral counting found later -- were the source of the fit to the process (\ref{npcapture}) obtained by Riska and Brown~\cite{riska-brown}. What was surprising at the time was that these exchange currents derived in the soft-pion limit worked well also at momentum transfers as high as $q^2 \sim 15$ fm$^{-2}$ in the electrodisintegration of the deuteron~\cite{hockertetal}
\be
e+d\rightarrow e^\prime + p + n.\label{ed}
\ee
It was not clear then why shorter-distance contributions, namely, the exchange of heavier mesons for which no systematic tools were available then, were not needed. The understanding of this result required an input from chiral perturbation theory. This story was described already in 1981 in a comment article by Gerry and me~\cite{BR-comments}.

As far as I know, the first application of the so-called chiral perturbation theory (ChPT for short) to nuclear processes was made in a 1981 lecture in Erice~\cite{MR-erice81}\footnote{At this school, I suggested -- and discussed how -- to implement chiral counting in computing pion-exchange currents.  It was, however, incomplete then but later completed as described below. Ulf Meissner was one of the graduate students in that Erice school. He probably does not recall that lecture; he has since then made significant contributions to the field of applying chiral perturbation theory to the physics of pion-nucleon systems and nuclear effective field theory.}. What motivated me to resort to chiral perturbation theory was the article by Weinberg where his ``folk theorem" on effective field theory was proposed~\cite{weinberg-folk}. The idea was that the soft-pion theorems were just the leading order terms in a field theory that allows one to calculate to all orders of low-energy processes involving pions in chiral perturbation theory. This meant that one should be able to compute systematically the full exchange currents going beyond the soft-pion theorems as  Weinberg did for pion interactions. Nucleons were involved in nuclear processes in addition to the pions considered by Weinberg. However, if one were to confine to two-nucleon processes, then one simply could take over Weinberg's counting rule -- with some tractable additional terms such as $\Delta$ -- to the exchange current problem, as I did in the Erice lecture. The counting associated with the nucleons (internal and external) as well as others did not affect -- apart from an overall factor -- the power counting to the order discussed then. Later Weinberg wrote down the complete counting rule for nuclear processes that includes both nucleons and pions~\cite{weinberg91}, which subsequently led to a more precise formulation of chiral perturbation approach to the exchange currents~\cite{MR91}. The power of chiral perturbation treatment was that it explained in a simple way how the soft-pion terms dominate the corrections in certain processes, while allowing to calculate higher-order terms.

The modern state-of-art calculations following the above strategy account for the process (\ref{npcapture}) within 1\%~\cite{parkPRL} and (\ref{ed}) within a few \% accuracy.

The same current algebra argument made an equally clear-cut prediction on axial charge transitions in nuclei~\cite{KDR}. The pertinent observation was that the nuclear beta-decay transition from nucleus $A$ to $A^\prime$
\be
A(J^\pm)\rightarrow A^\prime (J^\mp)+ e^+ (e^-) +\nu (\bar{\nu}),\ \ \Delta T=1\label{axialcharge}
\ee
where $J$ and $T$ are, respectively, the angular momentum and isospin and the superscript $\pm$ stands for the parity, is predominantly given by the time component of the axial current and that there the impulse approximation contribution is strongly suppressed~\cite{warburton}. Chiral perturbation theory straightforwardly explains why the two-body exchange axial-charge operator is also dominated by the soft-pion terms. This observation turns out to have a nontrivial consequence on how chiral symmetry manifests itself in dense medium~\cite{KR91}.

Looked from the present vantage point of effective field theory of nuclear physics, what I have described here is not strictly speaking consistent. The state-of-art wave functions are obtained with a sophisticated phenomenological potential (SPP) that we may consider to be of high order in chiral perturbation theory, say, $\nu^p$, with $p\gg 1$. What this means is that the higher one goes in chiral counting in ChPT, the closer the potential comes closer to the SPP. However, the current is computed at some finite order $\nu^q$ with $q\ll p$. This necessarily entails a mismatch in chiral counting between the potential and the current. The mismatch is a practical one in that it can be avoided if one can go to high enough an order in chiral expansion. It is not the sort of mismatch that is present in the Weinberg scheme in which there is a separation between the ``irreducible" graphs -- calculated to some finite order -- defining the potential and the ``reducible" graphs that enter in solving the Schr\"odinger equation or Lippmann-Schwinger equation with the given potential entailing an infinite order~\cite{KSW}.  Our philosophy here is that this strict order-by-order adherence to full consistency in chiral counting is not necessary for attaining the accuracy desired, given that the state-of-art wave functions are available. The power of this approach is that it can be applied not only to few-nucleon systems but also to heavy nuclei for which reliable wave functions are available such as in Pb nuclei~\cite{warburton}. This point was stressed in \cite{MR-prediction}.

This strategy has proven to be successful in accurately describing responses to external fields in a variety of other processes. This matter will be described elsewhere in this volume~\cite{KR-GEBfest}, so let me just give a list of what has been accomplished. In addition to the radiative $np$ capture and the deuteron electrodisintegration, calculations of unprecedented accuracy have been performed for, e.g., (a) the solar $p+p\rightarrow d+e^+ +\nu_e$~\cite{parketal}, (b) the solar $hep$\ ($^3$He $+p\rightarrow ^4$He + $e^+ +\nu_e$) process~\cite{parketal,tsp-kk-hep}, (c) the $hen$\ ($^3$He $+n\rightarrow ^4$He + $\gamma$) process~\cite{park-song,marcucci-hen}, and (d) the muon capture on deuteron and $^3$He~\cite{marcucci-mucapture}.

\section{The Little (Chiral) Bag and the Cheshire Cat}
In the development described so far in addressing nuclear interactions, there was no reference to the microscopic degrees of freedom of QCD, namely, quarks and gluons. The natural question raised is how the microscopic picture, QCD, maps to the macroscopic picture, nuclear EFT. This question was addressed in 1978.

In Fall 1978, I arrived at Stony Brook for an academic year. It was the second in my regular one-year visits to Stony Brook to work with Gerry -- which lasted for many years afterwards. Returning from his visit to Caltech, Gerry suggested to look at how one could make the ``bags" in which three quarks are confined smaller in nuclear environment. In Caltech, Gerry had conversations with Richard Feynman in which the size of the MIT bag was discussed. At the time, the MIT bag model was very popular. In this model, hadrons are described as three quarks for baryons and a quark and an anti-quark for mesons confined within a bag of size of typically $\sim 1$ fm in radius. To our thinking, this size for the nucleon was too big. There were also indications from high-energy proton-proton scattering as well as experiments on high mass muon pairs for a nucleon with core size of $\sim 0.2 - 0.3$ fm~\cite{orear}. There are more recent indications of similar size in deep inelastic scattering off nucleon~\cite{petronzio}. To us, a nucleon of the bag size $\sim 1$ fm in a heavy nucleus, say, Pb nucleus, was like a ``grapefruit" in a salad bowl. There was something basically bizarre about this. The puzzle that we were having was how can the picture of rigid balls of $\sim 1$ fm radius substantially overlapping each other in the bag model in, say, Pb where the average inter-nucleon distance is $\sim 2.2$ fm be reconciled with the highly successful shell model or more specifically for us, the effective field theory description given above where point-like nucleons interact via the exchange of ``soft" mesons -- which is essentially the old ``standard nuclear physics approach" (SNPA for short) anchored on Yukawa theory.

We proposed in that year that the resolution of this puzzle was {\it in the way chiral symmetry is manifested} in the structure of the nucleon~\cite{BRlittlebag}. With only the MIT bags in many-nucleon systems, pions -- accepted nowadays as the key ingredient of the spontaneously broken chiral symmetry of QCD -- were missing. Pions had to be introduced in consistency with QCD. It was then recognized that pions could be brought in by imposing the conservation of the baryon axial current~\cite{chodos-thorne}. What we chose to do was that the pion field be coupled to the quarks at the bag boundary and do not penetrate into the bag. This procedure adopted by Callan, Dashen and Gross~\cite{CDG} was natural in that the interior of the bag was supposed to be in the Wigner-Weyl (Wigner for short) mode whereas the exterior was supposed to be in the Nambu-Goldstone (Goldstone for short) mode. This structure turns out to possess an intricate feature known as ``the Cheshire Cat phenomenon." It may be taken as a particular {\it manifestation of chiral symmetry} in the guise of ``quark-baryon continuity." I will have more to say on this below.

The main idea of \cite{BRlittlebag} was as follows. Suppose that the bag were squeezed by an external pressure to a small size. Then the energy density of the confined quarks would increase making the quark condensate $\la\bar{q}q\ra$ -- which is non-zero in the vacuum -- go to zero, as expected in the Wigner phase. With the radius squeezed to a small size, $\sim 0.3$ fm,  the energy of the quarks would increase accordingly $\propto 1/R$, where $R$ is the bag radius, due to the kinetic energy of the confined quarks. The assumption was that the self-energy effect due to the coupling to the pion outside would provide a large enough attraction to cancel the increase in mass.

Considerations of subtle quantum effects in the chiral bag structure indicated, however, that this picture was not correct in its original form. Apart from the fact that such a strong attractive self-energy effect was not obvious, there was a problem associated with an anomaly -- a quantum effect -- in the baryon number $U(1)$ current at the boundary which was not taken into account in the ``little bag" scenario. While the color is confined within the bag, the baryon charge which is conserved classically is not, at the quantum level. The quantum anomaly generated by an induced axial gauge field arising from the continuity of the axial current at the boundary makes the baryon charge leak out into the outside of the bag. The pion field lodged outside has the mean field of the hedgehog form $\pi=\frac 12{{\tau}}^a \pi^a(\mathbf{r})\sim  \frac{f_\pi}{2}{\vec{\tau}}\cdot\hat{\mathbf{r}}{\rm cos}\theta$ where $\theta$ is known as the ``chiral angle" and goes from 0 to $\pi$ as a function of radius. This object has a topology and picks up the leaking baryon charge. It was first shown in \cite{magic-angle} that for the chiral angle $\theta=\pi/2$ (dubbed as ``magic angle"), the baryon charge is partitioned equally between the inside and outside of the bag, and then proven generally in \cite{goldstone-jaffe} that the partition takes place for {\em any} angle. This led to the suggestion that physics should not depend on the bag radius, a phenomenon given the name ``Cheshire Cat"~\cite{CC}. This phenomenon can be rigorously established in (1+1) dimensions thanks to the exact bosonization technique. However, bosonization technique for (3+1) dimensions was -- and still is -- not available and hence the phenomenon could only be approximate. Highly approximative though it may be, the Cheshire Cat notion has met with success for a variety of nuclear phenomena studied in the model~\cite{CC-MR}. This picture has reappeared in a holographic dual QCD description of the baryon as described by Ismail Zahed in this volume~\cite{IZ}.

The lesson from the Cheshire Cat phenomenon is that for long-wavelength probes, the bag could be shrunk to zero. When the bag is shrunk to zero, the resulting baryon is the skyrmion soliton~\cite{skyrme}, with the confinement size rendered unphysical.

What happens here illustrates even a subtler way chiral symmetry can be manifested in {\em nuclear} systems.

A similar analysis showed that the color could also leak out due to the well-known $U(1)_A$ anomaly~\cite{colorleakage}. In contrast to the baryon charge where the hedgehog pion absorbs the baryon charge, there is nothing  outside of the bag that can absorb the leaking color charge. This required that the color-leakage be stopped by fiat, that is, by putting by hand a boundary condition in the Lagrangian. This implied that the color symmetry was broken at the classical level and restored at the quantum level. This procedure in some aspect resembles the Green-Schwarz anomaly cancelation mechanism in string theory. The resulting chiral bag Lagrangian was found to reproduce the Witten-Veneziano formula for the $\eta^\prime$ mass, i.e., the resolution of the $U(1)_A$ anomaly problem~\cite{etaprime}.

An interesting modern development is that the Little Bag structure seems to appear in a strong coupling lattice (SCL) QCD calculation of nuclear matter~\cite{deforcrand}\footnote{It is interesting -- and perhaps significant -- that a meson-exchange picture arises in this SCLQCD calculation in a manner resembling the old Heitler-London model of Cutkosky~\cite{cutkosky}.}, so it appears that somehow the small size core~\cite{BRlittlebag,orear,petronzio} refuses to disappear in the Cheshire Cat smile. I will come back to this matter when I describe the instanton structure of the nucleon in holographic dual QCD.

\section{B(rown)R(ho) scaling}
In Fall 1990, I was spending a semester at Seoul National University giving a special topics course. Early in my stay at SNU, I received a fax from Gerry that contained a copy of the draft of Weinberg's paper on chiral Lagrangian approach to nuclear forces -- published later in \cite{weinberg91} -- and a summary of his thoughts on Weinberg's approach concerning the role of the $\rho$ meson in nuclear physics. Gerry mentioned in the fax that he was disagreeing with Weinberg on the necessity of having the $\rho$ meson explicitly in the construction of the nuclear forces. In Weinberg's chiral perturbation approach, the $\rho$ meson (and all other massive degrees of freedom than the nucleon) have been integrated out, with their effects, if any, appearing in higher chiral-order terms. Gerry was stressing the need of having the $\rho$ {\em explicitly} because of what we found in our work on the tensor forces in nuclei where the cancelation between the pion tensor and the $\rho$ tensor -- which come with an opposite sign~\cite{BRtensor} -- became important. The argument in \cite{BRtensor} was that if one assumed that the ``intrinsic" hadron masses dropped in increasing density as naive constituent quark model indicated, then the cancelation in the tensor forces would become greater as density increased and Gerry was arguing that such an effect -- that depends directly on the $\rho$ degree of freedom -- would be hard to capture if the $\rho$ meson were absent in the consideration.

This fax from Gerry led to two notable developments on my side, one with and the other without Gerry. The one without Gerry was that Weinberg's (draft) paper immediately -- and almost trivially -- indicated how to complete the (incomplete) formulation given in my 1981 Erice lecture of exchange currents in chiral perturbation theory~\cite{MR91}. The other with Gerry was the idea of in-medium scaling based on chiral Lagrangian, i.e., the BR scaling~\cite{BRscaling}. The former cleared my fuzzy understanding of the role of the soft-pion theorems in certain nuclear electroweak processes and paved the road to my part in the work described in the contribution of Kuniharu Kubodera and myself~\cite{KR-GEBfest}. The latter put on a more formal ground the argument we developed in \cite{BRtensor} for the in-medium weakening of the tensor forces. The special role of vector mesons in nuclear interactions will become more prominent in a modern holographic dual QCD model of the nucleon mentioned below.

The idea behind the BR scaling went as follows.

The Cheshire Cat principle suggested that baryons in nuclear matter could be approximated by skyrmions considered as chiral bags shrunk to a point. Since the skyrmion has well-defined charge (electric, baryonic etc.) distributions, it should be easy to treat its property in medium. Here the first question was: What is the appropriate Lagrangian for the skyrmion description? The second question was: Given such a Lagrangian, how does one treat nucleon properties in medium?

In addressing the first question, suggested by Weinberg's folk theorem, we needed to identify the degrees of freedom relevant to the low-energy scale involved. Then guided by the large $N_c$ arguments~\cite{largeNC}, one would write an effective theory for QCD at low energy as a power series (in derivatives and the pion mass) in terms of the chosen degrees of freedom. For simplicity, we took the simplest Lagrangian with Goldstone pions only with the heavy degrees freedom such as vector mesons, scalar mesons, glueballs etc. assumed to be integrated out. Baryons should arise as topological solitons from such a Lagrangian, though strictly valid in the large $N_c$ limit. At near zero energy, we knew what it should look like from the current algebras,
\be
{\cal L}_{\bf CA}=\frac{f_\pi^2}{4}{\tr}(U^{-1}\del_\mu U)^2.\label{CA}
\ee
By itself, this term does not support a soliton, so to get a stable soliton, one needs higher derivative terms. Skyrme in his path-breaking paper~\cite{skyrme} put by hand a quartic term -- called Skyrme term -- of the form
\be
{\cal L}_{\rm Skyrme}=\frac{1}{32e^2}{\tr} [U^{-1}\del_\mu U,U^{-1}\del_\nu U]^2.\label{skyrmeterm}
\ee
This term does stabilize the soliton. The Lagrangian consisting of the two terms (\ref{CA}) and (\ref{skyrmeterm}), called ``Skyrme Lagrangian" in the literature, has been extensively studied for the structure of baryons~\cite{zahed-brown} as well as finite nuclei and infinite nuclear matter~\cite{CNDII}. It is, however, too simple to be a realistic model for the baryon, even more so for many-nucleon systems, although it embodies the basic concept of solitonic baryon. There were lots of objections raised against the Lagrangian, among which most serious was that there is no reason why the derivative expansion should stop at the quartic order. We ignored all these caveats at the time. Some of these objections were lifted later when vector mesons, in fact an infinite tower of them in holographic QCD~\footnote{In the holographic dual QCD model to which I will return, the low-energy action in the large $N_c$ limit is given by a 5-D $U(N_f)$ (where $N_f$ is the number of flavors) Yang-Mills action $\propto (F_{AB})^2$ in curved space (ignoring higher order terms such as $(F_{AB})^3$). When all the vector fields that emerge in dimensional reduction to 4D \`a la Klein-Kaluza are set to equal to zero, one is left with precisely the Skyrme Lagrangian, with no other quartic terms. There is only one quartic term here, i.e., the Skyrme-term type. This is because the Skyrme term is the only quartic term with one time derivative coming from the reduction of the $\propto (F_{AB})^2$. It is worth noting that in the hQCD model, the Skyrme term has no role whatsoever in stabilizing the soliton (instanton). It is the Chern-Simons term involving the $\omega$ field that prevents the collapse of the soliton.}, were incorporated.

In \cite{BRscaling}, we decided to bypass dynamical contents but focus only on, or exploit, the chiral symmetry structure of the Skyrme Lagrangian for describing the properties of hadrons in dense medium. The immediate goal here was to justify the scaling behavior of the hadrons used in \cite{BRtensor}. For this, we had to figure out how to do an effective field theory calculation with the Skyrme Lagrangian in a dense medium. We were principally motivated by the success achieved by Walecka mean field theory for nuclear matter~\cite{walecka}. This is anchored on the premise that such a mean-field approach corresponds to doing Fermi-liquid theory and that such a mean-field theory can be constructed in a chirally symmetric way as will be elaborated below. Therefore the strategy was to write a chiral Lagrangian in the form of the Skyrme Lagrangian and then do the mean field. For this, we felt that a scalar meson of the mass comparable to that used in Walecka model would be needed to address many-nucleon matter. The question was: What is this scalar meson? It was surely not the scalar meson $\sigma$ in the linear $\sigma$ model, because we were dealing with the nonlinear sigma model, for which the $\sigma$ of the linear $\sigma$ model, being heavy with mass $\gsim 1$ GeV, has been integrated out. The scalar needed in Walecka model had to be a chiral singlet with mass $\lsim 600$ MeV. In \cite{BRscaling} the scalar was implemented with a dilaton field $\chi$ that we assumed to interpolate only the soft -- and not hard -- component of the trace anomaly of QCD that is locked to the quark condensate, the order parameter of chiral symmetry~\footnote{The trace anomaly is given by $\theta_\mu^\mu=\del_\mu D^\mu$ where $D_\mu$ is the dilatation current. In QCD, $\theta_\mu^\mu=\frac{\beta(g)}{2g}(G_{\mu\nu})^2$ in the chiral limit with vanishing quark mass. Defining the dilaton field $\chi$ as interpolating the gluon condensate, it is useful to decompose, following \cite{miransky}, $(G_{\mu\nu})^2\sim (\chi^4)_{\rm soft} + (\chi^4)_{\rm hard}$ and associate the soft component tied to chiral symmetry breaking. This subtle point in doing the separation was clarified much later in a more satisfactory way in collaboration with Hyun Kyu Lee in the WCU program at Hanyang University~\cite{LR09}. The soft dilaton will be denoted simply as $\chi$ in what follows.}. At the time, quite a few particle theorists were interested in connecting conformal symmetry, more specifically, scale symmetry to chiral symmetry and we adopted their strategy. The idea is to make the action for the Goldstone bosons and the vector mesons -- apart from the chiral symmetry-breaking mass term -- scale-invariant and attribute the scale non-invariance, i.e., the trace anomaly, to a potential $V(\chi)$~\cite{miransky,EL}. This meant that one should multiply the Lagrangian ${\cal L}_{\bf CA}$ by $(\chi/f_\chi)^2$ where $f_\chi$ is the vev of $\chi$, $f_\chi=\la\chi\ra_0$, and leave ${\cal L}_{\rm Skyrme}$ (the action of which is scale-invariant) untouched. Similarly for the action with vector mesons. Such an action would lead to $\theta_\mu^\mu\sim \chi^4$ without involving other fields except for the pion mass term.

The next step is then to consider the soliton solution of the resulting Lagrangian as giving the baryon and construct an effective Lagrangian that has as the pertinent degrees of freedom the pion, the vector mesons, the scalar $\chi$ and the nucleons~\footnote{The Skyrme Lagrangian does not have the vector mesons but we can recover their properties from the hidden local symmetry Lagrangian~\cite{bandoetal,HY:PR} from which it can be considered to come from. This is simply because the two Lagrangians are ``gauge-equivalent."}. Taking the mean field \`a la Walecka, the Lagrangian will then give nuclear matter in which the in-medium vev $\la\chi\ra^*$ will figure as a parameter of the Lagrangian. This led to the scaling law of \cite{BRscaling}
\be
\Phi(n)=m_N^*/m_N\approx m_V^*/m_V\approx m_\chi^*/m_\chi\approx f_\pi^*/f_\pi\label{BR}
\ee
where $f_\pi^*\approx f_\pi (\la\chi\ra^*/f_\chi$) and the subscript $V$ stands for the vector mesons $\rho$ and $\omega$. For this relation to make sense, it is crucial that the $\chi$ field be associated with the soft component of the gluon condensate, the hard component remaining unchanged as density increases. At the time this relation was proposed, there was a big misunderstanding because people took $\chi$ to have both components as it appears in the trace anomaly.
\subsection{``Deriving" BR scaling}
There are several different ways of interpreting, or arriving at, (\ref{BR}). Let me discuss a few here.

Chiral symmetry enters in (\ref{BR}) through the {\em locking} with the soft scalar condensate and hence manifests itself in the mean-field as formulated. Clearly the connection to QCD is not direct. Thus it is not clear how that relation is connected to QCD-motivated quantities such as in-medium QCD sum rules. It has been suggested that a Walecka-type mean field theory for nuclear matter is equivalent to Landau Fermi liquid fixed point theory~\cite{matsui}. Now assuming this equivalence, the relation (\ref{BR}) holding at nuclear matter density $n_0$ can be considered as a fixed-point relation valid at $n=n_0$. In fact such a BR-scaling Lagrangian at mean field is found to give a correct description of nuclear matter~\cite{songetal}. Now the density dependence of the parameters such as mass (\ref{BR}) and coupling constants simulate part of the higher-dimension operator terms added to the Walecka-type mean-field models in order to make good fits to finite nuclei as well as infinite matter~\cite{ring}. This could be interpreted as (a part of) the rearrangement terms brought in through many-body (typically, three-body) forces.  The scaling parameter $\Phi(n=n_0)$ can be related to the Landau parameter $F_1$ at the fixed point through the nuclear anomalous orbital gyromagnetic ratio~\cite{friman-rho} as we will see below. Gerry understood this connection before he started working on BR scaling in dilepton processes in heavy-ion collisions. This issue became later highly controversial with no paucity of misunderstandings.

The fact that the scaling factor (\ref{BR}) is tied to the Fermi liquid fixed-point parameter which encodes a part of quasiparticle interactions clearly warns of the subtlety in the way chiral symmetry can be manifested in many-nucleon systems. As it stands, it encompasses both the quasiparticle nature of mesons and baryons and their interactions. Obviously the absence of visible effects does not necessarily signify the absence of chiral symmetry effects. A good example of this is given by the {\em invisible} effect of (\ref{BR}) in the Dirac phenomenology that I will mention below. And a similar story can be told of the JLab result as I will explain.

If one is probing a system away from the fixed point as in the case of heavy-ion processes, the question arises in what way the relation (\ref{BR}) can figure in the process. In particular, one would like to know how the scaling works near the chiral phase transition.

The first possibility is to consider the pion decay constant $f_\pi$ as an order parameter of chiral symmetry, presumably locked to the quark condensate $\la\bar{q}q\ra$. In this case, if one were to naively extrapolate (\ref{BR}) all the way to the chiral phase transition point at which $\la\bar{q}q\ra=0$ (in the chiral limit), then all the masses would go to zero. However, this is probably too naive or even wrong. In hidden local symmetry theory where the nonlinear sigma model is elevated to a local gauge theory~\cite{bandoetal}, it was shown by renormalization group at one-loop order~\cite{HY:PR} that when the quark condensate goes to zero, independently of how it is driven, the hidden gauge coupling $g$ (which is related to the vector coupling $g_{\rho\pi\pi}$ or $g_{\rho NN}$) goes to zero, arriving at what is called ``vector manifestation (VM) fixed point." There it is the hidden gauge coupling $g$ that is the order parameter, not the pion decay constant. Now since the vector meson mass goes as $m_V^2\sim g^2$, the vector meson mass must scale near the VM fixed point as~\footnote{Gerry had been arguing that the $\rho$ mass was the order parameter long before the VM was discovered. In fact, the VM vindicated Gerry's point.}
\be
m_\rho^*/m_\rho\simeq g^*/g\simeq \la\bar{q}q\ra^*/\la\bar{q}q\ra\rightarrow 0.\label{VM}
\ee
This is a clear prediction {within the HLS theory framework.} It is not known what happens if one includes more vector mesons and/or other degrees of freedom. There is an analysis with constituent quarks in addition to the pions and the vector mesons, which suggests that the constituent quark mass $m_Q$ {\em also} scales very near the VM fixed point as $m_Q^*/m_Q \simeq g^*/g\simeq \la\bar{q}q\ra^*/\la\bar{q}q\ra\rightarrow 0$.

The situation is even less clear when there are an infinite tower of vector mesons as implied by holographic QCD models.

I should point out that (\ref{VM}) is the only rigorous demonstration known so far of what happens to the meson mass at the chiral transition, although it can be established strictly within the specified framework. I know of no other statement on the property of the meson mass that can be taken as rigorous. Lattice QCD cannot offer anything yet on it either in temperature or in density. No other models that I know of do the RG analysis in a consistent way. Most of the models are hodge-podge mixtures and can hardly be trusted.

That this happens at high temperature was shown within the HLS theory without introducing fermions. In density, however, fermions are needed. One might try to generate baryons as skyrmions from the HLS mesonic Lagrangian, but one could work equally well with quasiquarks (or constituent quarks). There one can show that the vector meson mass does also go to zero at high density in a manner analogous to temperature. Considered in terms of the constituent quarks mentioned below, the nucleon, a bound state of three constituent quarks, is expected to become massless at the chiral transition point. So very near the transition point, the BR scaling (\ref{BR}) seems to be recovered with $\Phi$ scaling linearly in $\la\bar{q}q\ra$. Unfortunately, up to date HLS can say nothing definite away both from the vacuum and from the VM fixed point.

There is at present no unquenched lattice data which supports or refutes (\ref{BR}). However, there is an evidence for the relation (\ref{BR}) in strong coupling lattice QCD in the leading order in $1/g_c$ and $1/d$ where $g_c$ is the color coupling and $d$ is the space dimension~\cite{ohnish}. It remains to be seen what happens when $1/g_c$ and $1/d$ corrections are included.

The simplest -- and perhaps the most appealing -- way to get at (\ref{BR}) is the constituent quark model. It leads to a similar relation as that from the skyrmion picture but in a much simpler way. A recent discussion on this approach is based on the Manohar-Georgi model~\cite{manohar-georgi} which was recently given a strong backing by Weinberg in the large $N_c$ limit~\cite{weinberg10}. In this model, the relevant degrees of freedom are the constituent quarks, the pions and the gluons. They figure in the phase where chiral symmetry is broken, but quarks and gluons are deconfined. The essential idea in this work is that the constituent quark mass, which is of order of the renormalization scale $\Lambda_{QCD}$, is $N_c$-independent. If one supposes that the constituent quark mass comes from the spontaneous breaking of chiral symmetry -- which is, however, not implied by the model itself, then it seems reasonable to think that the constituent quark mass will go to zero when the quark condensate goes to zero (in the chiral limit) in a manner analogous to the VM of hidden local symmetry. In the Manohar-Georgi model, the nucleon mass scales as $m_N\sim \la\bar{q}q\ra$. Now if the vector mesons and baryons are loosely bound states of constituent quarks and antiquarks, then as density increases, the universal scaling of the type (\ref{BR}) will follow with $f_\pi^*/f_\pi$ replaced by $(\la\bar{q}q\ra^*/\la\bar{q}q\ra)^q$ where $q$ is some positive power $\sim 1/2-1$~\footnote{Some models like NJL or QCD sum rules give $q=1/2$, whereas HLS theory gives $q=1$.}.

\subsection{``Seeing" BR scaling}
\subsubsection{Nuclei and nuclear matter}
Up to nuclear matter density $n_0$ to which ample experimental data are available, the relation (\ref{BR}) works fairly well. I know of no cases where the scaling (\ref{BR}) is at odds with nature~\footnote{Since this matter is outside of the context of this note, I put it in a footnote. I know of one case in which the effort to {\em directly see} the scaling (\ref{BR}) with the $\rho$ meson in nuclear medium has ``failed," i.e., the CLAS/JLab~\cite{JLab}. Given the background analysis and other details of experimental procedure, one should be cautious of the interpretation of the result. Let me assume for the sake of discussions that one is given an experimentally determined spectral function. Since one is dealing with a low density no greater than that of the nuclear matter, one can approach the spectral function in two ways. (1) Start with a theory with the parameters of the effective chiral Lagrangian defined at the matter-free vacuum and compute the spectral function in chiral perturbation theory to as high an order as feasible. Here the vacuum parameters, not the scaling parameters (\ref{BR}), should be used. (2) Start with a chiral Lagrangian with parameters defined at a ``sliding vacuum" with the scaling (\ref{BR}) and then compute corrections as in (1). One possibility is to start from the Fermi-liquid fixed point and treat deviations from the fixed point as was done in \cite{friman-rho}. These two ways should give the same result if done fully consistently. It is not unlikely that there are cancelations in the second approach which is not present in the first, as exemplified by the example of the Dirac phenomenology~\cite{Brown-Hintz} mentioned below. Gerry has emphasized a sort of hidden scale invariance in nuclear processes that can mask the scaling behavior as it does in the Dirac phenomenology. Furthermore, the usual vector dominance (\`a la Sakurai) -- adopted by most of the theorists -- is violated in dense medium (more rapidly than in hot medium mentioned below), so the interpretation becomes highly tricky, and it would be difficult to sort out all (theoretical) backgrounds well enough to un-mistakenly ``identify" the scaling (\ref{BR}) in this sort of experiments.\label{footnote}}. Let me list just a few notable ones, in which Gerry participated directly and indirectly:

\begin{itemize}
\item {\it Nuclear tensor forces}

The early suggestion~\cite{BRtensor} that the scaling (\ref{BR}), through increased cancelation between the pion and $\rho$ tensor forces mentioned above, would modify the structure of the tensor forces was reconfirmed in a striking way in the C14 dating beta decay process~\cite{Holt}. This is an exceptional case in nuclear physics where dialing certain parameters amplifies enormously the effect of the ``dropping" $\rho$-meson mass.
\item {\it Proton-nucleus scattering}

Among numerous applications of (\ref{BR}) to nuclear scattering processes, the one where Gerry actively participated was proton-nucleus scattering at several hundreds of MeV~\cite{Brown-Hintz}. Incorporating the scaling (\ref{BR}) removed certain persistent discrepancy in analyses with the nonrelativistic impulse approximation, a standard technique at the time, and improved agreement with experiment. There are other processes where similar improvements are obtained, that I will not go into.

\item {\it Anomalous orbital gyromagnetic ratio in nuclei ($\delta g_l$)}

Since BR-scaling chiral Lagrangian in the mean field can be identified with Landau-Migdal theory~\cite{friman-rho}, one can derive the Migdal formula $\delta g_l=\frac 16 \frac{F_1^\prime -F_1}{1+F_1/3}\tau_3$~\cite{migdal} in terms of $\Phi$ of Eq.~(\ref{BR}). The scaling parameter $\Phi$ enters in place of the Landau-Migdal parameters $F_1$ and $F_1^\prime$, giving $\delta g_l=\frac{4}{9}[1/\Phi -1 -\kappa_\pi]\tau_3$ where $\kappa_\pi$ is given by the pion, so is accurately known. This formula gives in the Pb region $\delta g_l^p\approx 0.22$ for $\Phi (n_0)\approx 0.8$ given by QCD sum rules. The result from experiment on giant dipole resonances is $0.23\pm 0.03$. This is a clear indication that the scaling, with an intricate, albeit indirect, link to chiral symmetry, is encapsulated also in the Landau parameters at the fixed point.

\item {\it Warburton ratio in heavy nuclei}

As mentioned above, the time component of the axial current in nuclei is kinematically suppressed, so the ``impulse approximation" for the axial charge transition (\ref{axialcharge}) does not dominate the transition matrix element. In fact, the soft-pion exchange current correction can be   rather big~\cite{KDR}. At just about the same time as the scaling relation (\ref{BR}) was proposed, Warburton found~\cite{warburton} that his analysis of the ratio (that I refer to as ``Warburton ratio") $\epsilon_{\rm MEC}=M_{\rm exp}/M_{\rm sp}$ where $M_{\rm exp}$ is the experimental matrix element of the axial charge operator and $M_{\rm sp}$ is the shell-model single-particle matrix element in the lead region, came out as $\epsilon_{\rm MEC}=2.01\pm 0.05$, considerably larger than the prediction $\sim 1.4$ based on the operator of \cite{KDR}. This enhancement was quantitatively explained by the scaling (\ref{BR})~\cite{KR91}. In fact the enhancement in $\epsilon_{\rm MEC}$ ranging from $A=12$ to $A=208$ observed in experiments is fairly well -- if not entirely -- explained in an extremely simple way by the scaling (\ref{BR})~\cite{BR:PR02}.

\item {\it Nuclear matter properties}

One of the subtlest aspect of chiral symmetry manifestation is that it can be made a part of nuclear many-body interactions. When formulated in terms of the sliding vacuum as in the cases given above and further elaborated below in Section \ref{DD}, the scaling (\ref{BR}) is embedded in the Landau parameters. Thus nuclear matter can be described in an economical way in a generalized mean-field formulation in terms of a chiral Lagrangian with (\ref{BR})~\cite{songetal}, which corresponds to the ``single-decimation" scheme mentioned below in Sect. \ref{DD}. It is true that such a formulation is not yet made quantitatively accurate, but it seems highly feasible to make it so if there are enough data to fit the parameters of the effective chiral Lagrangian that one starts with. I should also mention what is called ``double-decimation" scheme which  belongs to the same class of approaches for nuclear matter. This will be discussed in Section \ref{DD}.
\end{itemize}

The effects I have discussed thus far are all {\em indirect} in that the particles involved are off-shell. They are in fact closely connected with the meson-exchange currents that were clarified by effective chiral field theory as I described above. In this sense, one can only say that (\ref{BR}) is  not inconsistent with what has been observed in  Nature: They are clearly not a ``snap-shot" of an on-shell particle propagating with the scaling mass, which is what one wants. I will turn to this issue next. But before doing so, one can, however, ask whether there are no evidences that (\ref{BR}) contradicts Nature, in other words, is falsified by experiments. If the degrees of freedom figuring in a particular process have masses different from the free-space values with which physics has been done up to now, are there no cases where the agreement obtained without scaling masses is upset by the scaling masses?

This question was addressed by Gerry. One of the most prominent successes of the original (linear) Walecka mean-field model was the Dirac phenomenology on the spin observables in proton-nucleus scattering. One would naively think that the scaling of the type (\ref{BR}) would upset the most spectacular fit of the data. It turns out, however, that this was not the case~\cite{Brown-Hintz, BR:DD}: There was no inconsistency over the whole range of kinematics measured. Gerry has argued that there is a hidden scale invariance in nuclear many-body process that preserves certain spin observables. This is yet another illustration that {the absence of visible signals does not necessarily imply the absence of the scaling effects.}
\subsubsection{Dileptons}
Suppose a $\rho$ meson is produced inside a big chunk of dense matter and that one would like to ``see" how it behaves inside the dense matter. Now suppose that the detector is put inside the matter. It will then detect the $\rho$ meson propagating with its properties, such as its mass, couplings with other ``particles" etc. determined by the ``vacuum" affected by the density of the medium.

However, in all the experiments performed so far, the detector is outside of the medium. Given the situation, it was natural to think that the best means to take the ``snapshot" of the properties of the $\rho$ in medium would be to measure the spectral function of the dileptons produced via the process $\rho\to l^+ l^-$ (where $l=e, \mu$) inside the matter. The question asked was: Can the scaling (\ref{BR}) be singled out in the dilepton production in nuclear collisions when measured outside of the medium?

In order for the relation (\ref{BR}) to be applicable, the large $N_c$ approximation should be reliable. Recall that it is a mean-field relation and only in the large $N_c$ limit can it represent a quasiparticle. Whether or not the vector meson can be treated reliably as a quasiparticle with a ``narrow" width in density greater that of $n_0$ is of course totally unknown. If the residual (quasiparticle) interactions become too strong, then the quasiparticle picture will surely break down.

As emphasized, within the well-defined HLS scheme with the vector mesons $\rho$ and $\omega$ and the pion as the only degrees of freedom (with nucleons arising as solitons), the RG arguments do predict that the masses will go to zero according to (\ref{VM}) with the widths also going to zero. So in that scheme, as $\la\bar{q}q\ra\rightarrow 0$, there will be a sharp peak at low invariant mass. Will this be detected in a measurement with the detector outside of the medium?

Gerry applied the scaling (\ref{BR}) to the dilepton measurements in heavy-ion collisions in mid 1990's assuming that a similar scaling holds in temperature~\footnote{There is no strong theoretical argument to suggest that one can apply (\ref{BR}) to heavy-ion processes where high temperature is involved. Although,  unlike in density, there is lattice information on temperature, there are no unquenched lattice data for light-quark hadrons near the critical temperature. The only theoretical statement one can make is that the vector manifestation is {\em shared} by both temperature and density in the framework of HLS. But at present,  HLS can say little on what happens in between the vacuum and the transition point (both in $T$ and $n$).}, and it seemed to explain the low-mass dileptons seen in the SPS/CERN experiments~\cite{brown-ceres}. This scenario is called ``dropping mass scenario (DropMS)." It turned out that these dilepton data could also be accounted for if one were to take into account such standard nuclear many-body effects as particle-hole excitations (or ``sobars"), collisional broadening etc. Let me refer to this scenario where the strong mundane nuclear effects dominate as ``melting scenario (MeltS)." These many-body effects are expected to be there anyway even if one has the quasiparticle picture as the first approximation. The question would be whether the quasiparticle picture makes sense in the heavy ion process in question.  There have been long debates as to whether it is the DropMS {\em or} MeltS that is seen in the experiment. Gerry and I believe this is a wrong debate. The correct approach should be DropMS {\em and} MeltS together {\it treated consistently}, not DropMS {\em or} MeltS.

A succinct summary of both experimental and theoretical situations on dileptons is given in \cite{rapp}. The upshot can be put in one line: {\it The dilepton data in heavy-ion collisions can be more or less entirely explained by mundane (standard) nuclear interactions}. Does this mean that Gerry was wrong in \cite{brown-ceres}? My answer is: Not entirely. Before I explain this, let me mention that what the present interpretation in terms of the MeltS mechanism is telling us is that what has been seen in the experiments is what I would call the ``background" to what one is looking for as the {\em signal} for the working of chiral symmetry in dense/hot medium. That the background accounts for the observed result simply means that whatever is looked for has not been exposed. It does not necessarily mean that it is not there. It is currently said in some quarters that the ``melting mechanism" is a {signal for chiral symmetry restoration.} This is as absurd as saying that the explanation of the anomalous gyromagnetic ratio or absence of the deviation in the Dirac phenomenology mentioned above {\em evidences} chiral symmetry at work. The background, the ``gooey" stuff, could have been a consequence of just plain strong interactions regardless of whether there is chiral symmetry involved or not.

Now let me address the issue of whether or not one can rule out (\ref{BR}) or (\ref{VM}) from the dilepton results~\footnote{Both Gerry and I disown the mysterious green curve one sees in certain experimental papers and conference talks attributed to ``BR scaling" and invoked to rule out (\ref{BR}). Such a curve has absolutely nothing to do with what was originally predicted in \cite{BRscaling} or in later versions of the scaling.}. My point is that one cannot. The mechanism Gerry was advocating in \cite{brown-ceres} could very well be there but as we now know it should be accompanied by other intricate mechanisms contained in HLS formalism. For instance, his consideration did not take into account that there is ``hadronic free regime" in which the photon is most likely decoupled from the dropping-mass vector mesons. The RG consideration in HLS suggests that because of the scaling in the gauge coupling $g$ and the parameter $a$~\footnote{$a$ is a parameter in hidden local symmetry theory defined as the ratio $f_s^2/f_\pi^2$ where $f_s$ is the decay constant of the scalar that is Higgsed to give the $\rho$ the mass or the longitudinal component of the $\rho$. This parameter signals deviation from vector dominance in medium.}, in the temperature or density regime where (\ref{VM}) is applicable, the photon coupling to the BR-scaling vector mesons gets suppressed. There are several agents for this which individually may not be significant but become important together. One of them is that the naive Sakurai vector dominance picture breaks down in temperature and in density, so that the strength of the $\rho$ coupling to the photon gets suppressed. This can be exacerbated by the infinite tower effect seen in holographic QCD where the tower of vector mesons can play an important role in photon coupling~\cite{HMY}. This means that the dileptons seen in the measurement carry no unambiguous imprint, free of the background, of the vector mesons whose intrinsic mass has undergone the shift (\ref{BR})~\cite{blind}. In other words, even if the dropping-mass vector meson were there, it would be like a needle in the haystack swamped by the background of strong nuclear interactions. What is observed in the experiments is just the lepton pair coming from hadrons in on-shell processes taking place at the ``flash point"~\footnote{The ``flash point" in our language~\cite{hadronicfree} is defined to be the density or temperature at which particles with decreased masses and weak coupling flowing from the hadronic free regime recover most (not necessarily all) of their free-space masses and strong interactions. It is difficult to precisely pin down the flash density or temperature. But to give a rough idea, let me give some rough guesses here. In the case of density, there is an indication from dense matter simulated on crystals that it could lie at $\sim (1.3-2)n_0$ and in temperature, not far from the critical temperature $\sim 170$ MeV. Pinning down this point (either in density or in temperature) would be a challenge for the future.} and this process has little to do with the vacuum change that one is looking for. The same argument applies to the CLAS/JLab result on electro-produced $\rho$ mesons in nuclei. Somewhat detailed discussion on this issue was given above in the footnote \ref{footnote}.

In sum: What has been shown in this flurry of activities is simply that nucleons in nuclei do indeed interact strongly, which of course nobody would question. If this were all there is to it, then that would be the most boring situation. Maybe one is not doing the right thing in the search for the ``smoking gun"~\footnote{One cannot help but wonder, however, why in condensed matter physics, things could be equally complicated in a strongly-coupled system and yet there is beautiful order and simplicity as one gets to understand the system better, and phenomena with the same notions of quasiparticles and long-range order do not occur in the strong interactions.}. I should mention that the ambiguity and difficulty in extracting medium effects on the mass of the decaying particle as in dilepton processes, presumed to reflect the property of chiral symmetry in medium, was clearly pointed out by Yamazaki and Akaishi~\cite{yamazaki-akaishi}.
\section{Double decimation and $V_{lowk}$}\label{DD}

In Fall 1999, I was asked by Gerry to come to Stony Brook and give a series of lectures on renormalization group approach to Landau-Migdal theory of nuclear matter. At the time I was working on something related to what I did with Bengt Friman in 1996 at GSI on connecting BR scaling to the Landau parameters~\cite{friman-rho}. I was attempting to use Shankar's renormalization group approach to Landau Fermi liquid theory~\cite{shankar}, to identify the scaling at $n=n_0$ as the RG fixed point. The reasoning was that the chiral Lagrangian with BR scaling parameter, when treated in the mean field, corresponded to the Fermi liquid fixed point theory. My lecture was on this idea. What Gerry with his collaborators did was first to arrive at the $V_{lowk}$ in the first decimation~\cite{vlowk} and then go, via BR scaling, to the Fermi liquid fixed point using Babu-Brown induced interactions in the second decimation~\cite{brown:DD}. This was the idea of the double decimation~\cite{BR:DD}. What I did then was combining these two decimations into one. These two approaches more or less agree with each other, notably for $\delta g_l$ which was difficult to get right in the double-decimation method {\em without} BR scaling.

The double-decimation approach that implements BR scaling finds an additional support in the calculation of nuclear-matter properties in a standard many-body approach~\cite{ring-BR}. Here the second decimation was made using a ring-diagram technique in place of field-theoretic Fermi-liquid theory of \cite{brown:DD}. Again there is an equivalence between multi-body effects brought in by BR scaling and those found in empirical multi-body forces, e.g., of the Skyrme type.
\section{Strangeness in nuclear matter}
Kaon is massive because the  strange (current) quark mass is about 20 times the mass of the light quark. But it is lighter than other hadrons, such as the $\rho$. If one puts the kaons on the same footing as the pions and consider them as Goldstone particles with, however, the quark masses taken into account as corrections, then as first shown by Kaplan and Nelson~\cite{kaplan-nelson}, negatively charged kaons ($K^-$'s) can condense in nuclear matter at high density. This happens because the density, through attractive kaon-nuclear interactions, eats into the kaon mass, bringing the kaon mass to zero at some density. At this point, the kaon field picks up the vev $\la K\ra$ and the kaon Bose-condenses. Such a high density could be reached in neutron-star matter, so it was thought to be relevant to neutron stars. It was also thought to be reachable in heavy-ion collisions in the laboratories.

Together with Kuniharu who was on sabbatical at Stony Brook and Vesteinn (Thorsson), Gerry's student, we proposed a mechanism that we found was more pertinent for neutron stars~\cite{BTKR}. The scenario proposed by Kaplan and Nelson required that the pions condense first, which then trigger kaon condensation~\cite{politzer}. Our estimate showed that this scenario would take place at too high a density to be relevant to neutron stars. The reason was that pions were thought to be unlikely to condense at a density regime relevant to neutron stars due to highly nonlinear nuclear correlations at short distances. In the scenario that we came up with, the kaon mass need not fall all the way to zero before the kaon condenses. All that is required  is that the kaon mass $m_K$ drop at roughly the same rate as it does in the Kaplan-Nelson's at increasing density. The new ingredient in our scenario is to exploit that in neutron-star matter, there are electrons, and that the electron chemical potential $\mu_e$ increases as the matter density increases. If, at some density, the increasing $\mu_e$ crosses the dropping mass of the kaon, $m_K^* < m_K$, then the electrons will beta-decay into the kaons as
\be
e^-\rightarrow K^- + \nu_e.
\ee
These kaons will Bose-condense. The density at which the condensation takes place turns out to be not so high, say, around 3-4 times the nuclear matter density. Gerry, together with Hans Bethe and Chang-Hwan Lee,  made extensive studies of neutron star properties with this scenario, making some startling predictions on the maximum neutron star mass ($\sim 1.6 M_\odot$) and large number of small mass black holes in the Universe. This was summarized in a review article that he co-authored with Chang-Hwan Lee and me~\cite{BLR-stars}. We also speculated what the consequence of the increased number of black holes would be on cosmology~\cite{BLR-cns}.

There are several ways to provide support to the early, somewhat crude, prediction of the kaon condensation threshold. One possibility is to invoke the scaling relation (\ref{BR}) in the framework of Walecka mean-field theory starting from nuclear matter density, i.e., the Fermi-liquid fixed point~\cite{BR2walecka}. Another possibility is to exploit the sliding-vacuum method and start from the VM fixed point instead of the Fermi-liquid fixed point. This is in principle more astute since within the well-defined theoretical premise of hidden local symmetry, the VM fixed point is a robust starting point around which perturbation theory can be developed. Furthermore, the kaon condensation takes place nearer to the VM point than to the Fermi-liquid fixed point.  This was done in \cite{BLPR} obtaining the same result as approaching from the Fermi-liquid fixed point. These two approaches illustrate the point I raised above, namely, that the scaling (\ref{BR}) makes sense only in a framework where the starting point is defined with a sliding vacuum.
\section{The multifaceted skyrmion}
The dual character between the ``macroscopic" skyrmion picture and the ``microscopic" QCD picture for the baryon clearly points to a fuzzy notion of topology that enters in the former. The chiral bag model with the Cheshire Cat phenomenon attests to this ambivalence. There is a trade-in between topology and quark/gluon degrees of freedom. When the Skyrme model was resurrected in the early 1980's, a large number of nuclear and particle theorists had jumped on the ``bandwagon" and worked on it. However string theory re-emerged soon after with its ``first revolution," which took away most of the particle theorists involved in the Skyrme model into the main stream of string theory. That left the problem mostly to nuclear theorists who then went on to further elaborations of the model and to applying it to many-baryon systems. The output unfortunately was limited and even poor. The reasons for the paucity of significant results were multi-fold. First of all, as stressed throughout this note, the Skyrme model as it stands -- which was studied by most workers -- is at best only a poor caricature of QCD. Other degrees of freedom, notably the vector mesons --  possibly the infinite tower of them --, the scalar mesons etc. need be included. There is a clear signal for this need in the recent development of the holographic dual model of the baryon. (This matter is briefly touched on in the next section.) Secondly, even at the level of the pure Skryme model,  little is understood of what is contained in the soliton structure. Pure mathematicians and mathematical physicists, some as celebrated as Atiyah and Manton, are finding fascinating aspects of the model at the classical level, but nothing much is known at the quantum level. When the vector-meson degrees of freedom are included as clearly required by the Cheshire Cat principle of the chiral bag picture~\cite{IZ}, even less is understood. It is undoubtedly a difficult problem to resolve. For these and other reasons, the model was ``abandoned" by many workers in the field who then turned to more manageable and schematic models. The difficulty and the consequent abandonment were mistakenly identified by many of them as the defect in the soliton picture of the baryon itself, not as the failures or shortcomings of the workers themselves.

Gerry and I found this tendency of abandonment unjustified as well as disingenuous. We felt that what we have seen in the model for the baryon structure {\em per se} as well as for many-body systems, i.e., dense baryonic matter, was just a tip of a big iceberg and there were a wealth of hitherto unexplored fascinating phenomena to be uncovered. This led to our editing of the volume {\it The Multifaceted Skyrmion}~\cite{multifaceted}\ in which we gathered contributions ranging from nuclear physics to condensed matter physics to string theory, all dealing with the same notion of the skyrmion soliton ranging in dimensions from 3 to 5. The objective here was to illustrate by stunning developments in condensed matter physics and what is taking place in string theory how pervasive and rich the physics captured by the soliton structure can be. In fact, in strong interaction physics with the  intricacy involved with the Cheshire Cat phenomenon that is not shared by condensed matter, there is a growing evidence for certain phenomena associated with topology that are not visible in other models, such as for instance a half-skyrmion phase that appears at high density~\cite{half} that resembles the merons or half-skyrmions that are found in condensed matter systems~\cite{multifaceted}, e.g., bilayer quantum Hall phenomena, deconfined quantum critical point and thin chiral magnetic films.

\section{Recent developments}
I would like to end this note with an account of recent developments in which Gerry did not participate but that are closely related to both what he has done and what we have done together.

There has been a long-standing puzzle why Sakurai's vector dominance works well for the pion (and mesons in general) but fails famously for the nucleon. In \cite{BRW}, together with Wolfram Weise, we proposed to resolve this problem by means of a little chiral bag defined at the ``magic chiral angle" $\theta=\pi/2$ at which the baryon charge is shared equally inside and outside of the bag. In this case, the nucleon has a ``core" of $\sim 0.2-0.3$ fm as seemed to be required by the experimental observations mentioned above~\cite{orear,petronzio}. Recently this issue was revived by the discovery of an instanton picture for the baryon that arises from holographic dual QCD~\cite{SS}. The baryon in this theory essentially corresponds to a skyrmion built of the pion and an infinite tower of vector mesons~\cite{HRYY}. This picture turns out to give an extremely good description of the proton form factors in terms of a surprisingly simple formula derived from the infinite tower of the vector mesons~\cite{HR:ff}. The result, however, gives a structure of what corresponds to the ``core" that is drastically different from what was found before in terms of the quark-bag description~\cite{BRW} and also from the holographic Cheshire Cat~\cite{IZ}. I have at present no clear idea what this means in terms of QCD variables\footnote{It may be that ``integrating out" the high tower in the bulk (gravity) sector involving geometry gives a different physical picture from that of ``integrating out" higher energy scale degrees of freedom in the boundary (QCD) sector. The meaning of the ``core" in the QCD sector may not be directly mapped to what happens in the bulk sector.} and I am wondering what Gerry will say about it.

Another new development is a new form of scaling which modifies (\ref{BR}) for $n\gsim n_0$ that comes from the structure of cold dense baryonic matter simulated on crystal lattice~\cite{LPR}. When skyrmions are put on an FCC lattice to simulate dense matter, one finds that there is a phase change from the skyrmion matter to a half-skyrmion matter in CC. At a certain density $n_{1/2}$ above that of the nuclear matter -- the precise value of which cannot be pinned down at the moment, a skyrmion in a dense medium fractionizes into 2 half-skyrmions with the chiral symmetry putatively restored in the sense that ${\Tr}U\sim \la\bar{q}q\ra$ goes to zero at that density. When this happens, the nucleon mass stops dropping until it reaches the deconfined phase which takes place at a density $> n_{1/2}$ whereas the vector meson mass continues to drop with the dropping hidden gauge coupling $g^*$. Thus for $n\gsim n_{1/2}$ the scaling differs from (\ref{BR})
\be
m_N^*/m_N &\sim& {\rm const},\\
m_V^*/m_V &\sim& g^*/g.
\ee
This means that the tensor forces would look quite different below and above the density $n_{1/2}$, with a sudden change-over at that density. This would not affect the C14 dating process of \cite{Holt} described above, since up to $n_0$, it remains the same, but it makes a dramatic effect on the tensor forces at $n\gsim n_0$. Specifically the $\rho$ tensor is almost completely killed by $n\sim 3n_0$, leaving only the pion tensor effective. This is in strong contrast to what one expects with the scaling (\ref{BR}) which makes the whole tensor forces nearly vanish at that density. This would have a serious ramification on the symmetry energy which plays an essential role in the EOS of compact stars~\cite{LPR}. Again I wonder what Gerry will say about this.

\subsection*{Acknowledgments}
\indent\indent What I described in this note may not be exactly what Gerry sees as what has happened. I only hope that it does not deviate too much from his version. Gerry and I have written a large number of papers, many of which with provocative ideas, working with Sunday transatlantic phone calls whenever I was not at Stony Brook. It was a wonderful journey together, full of fun and excitement -- and never a dull moment -- and I would like to thank him for that.

This work was partially supported by the WCU project of Korean Ministry of Education, Science and Technology (R33-2008-000-10087-0).

\end{document}